\documentclass{svproc}
\usepackage{url}
\usepackage{graphicx}
\usepackage{pdflscape}
\usepackage{multicol}
\usepackage{float}
\usepackage{subfig}

\begin{document}
\mainmatter              
\title{Measurement of D-meson production in pp collisions with ALICE at the LHC}
\titlerunning{D-meson production with ALICE}  
%
\author{Ankita Sharma, for the ALICE Collaboration}
\authorrunning{A.Sharma} 
%

%
\institute{University of Jammu\\
\email{ankita.sharma@cern.ch}\\ 
}

\maketitle              

\begin{abstract}
The \textit{p}$\rm_{T}$-differential cross section of D mesons in the rapidity range  $|y|< 0.5$ was measured in pp collisions at $\sqrt{s}$ = 2.76, 7 and 8 TeV. D mesons were reconstructed in their hadronic decay channels by means of the invariant mass analysis. The D-meson production cross sections is compared among different energies and to pQCD calculations. Also the prompt D meson are studied as a function of the multiplicity of charged articles produced in inelastic pp collisions at a centre-of-mass energy of $\sqrt{s}$ = 7 TeV. The measurements are compared with model calculations.
\end{abstract}

\section{Introduction}
ALICE is the dedicated heavy-ion experiment at the Large Hadron Collider (LHC). Its main purpose is to investigate the properties of the deconfined state of strongly interacting matter produced in heavy-ion collisions, the Quark Gluon Plasma (QGP). The prime aim of the experiment is to study in detail the behaviour of nuclear matter at high densities and temperatures. The measurement of charm and beauty hadron production in Pb-Pb collisions is one of the main items of the ALICE physics program. Heavy-quarks are produced in hard scattering in the early stages of high energy nucleus-nucleus collisions and, due to their long life time, they are expected to be a powerful tool to investigate nuclear effects on charm production. The measurement of charm and beauty production in proton-nucleus collisions are important to understand cold nuclear matter effects on heavy-flavour production, and in proton-proton collisions these measurements are important for the comparison with perturbative Quantum Chromo-Dynamic (pQCD) calculations in a new energy domain and it provides a baseline to compare the results that are available in nuclear collisions.

\section{D-meson production cross section}
The measurement of charm and beauty production cross sections in pp collisions at the LHC constitutes an important test of pQCD calculations. D mesons were reconstructed in their hadronic decay channels by means of the invariant mass analysis~\cite{ref:ALICE7TeV}. The measurement of the production cross section of the charmed mesons $\rm D^{+}$ and  $\rm D^{*+}$ was done in pp collisons at $\sqrt{s}$ = 8 TeV, where D-mesons were reconstructed in the range 1 $<$ \textit{p}$\rm_{T}$ $<$ 24 GeV/\textit{c}. The prompt D-meson \textit{p}$\rm_{T}$-differential cross sections were compared with the FONLL theoretical predictions~\cite{ref:FONLL} as shown in Fig 1. It can be noted that, while fully compatible, FONLL predictions are on average lower than the measured cross sections. Therefore, FONLL tends to underestimate charm production in pp at 8 TeV, as was already noted at lower energies~\cite{ref:ALICE7TeV}. 
\begin{figure}[h!]
\begin{multicols}{2}
\includegraphics[width=5cm]{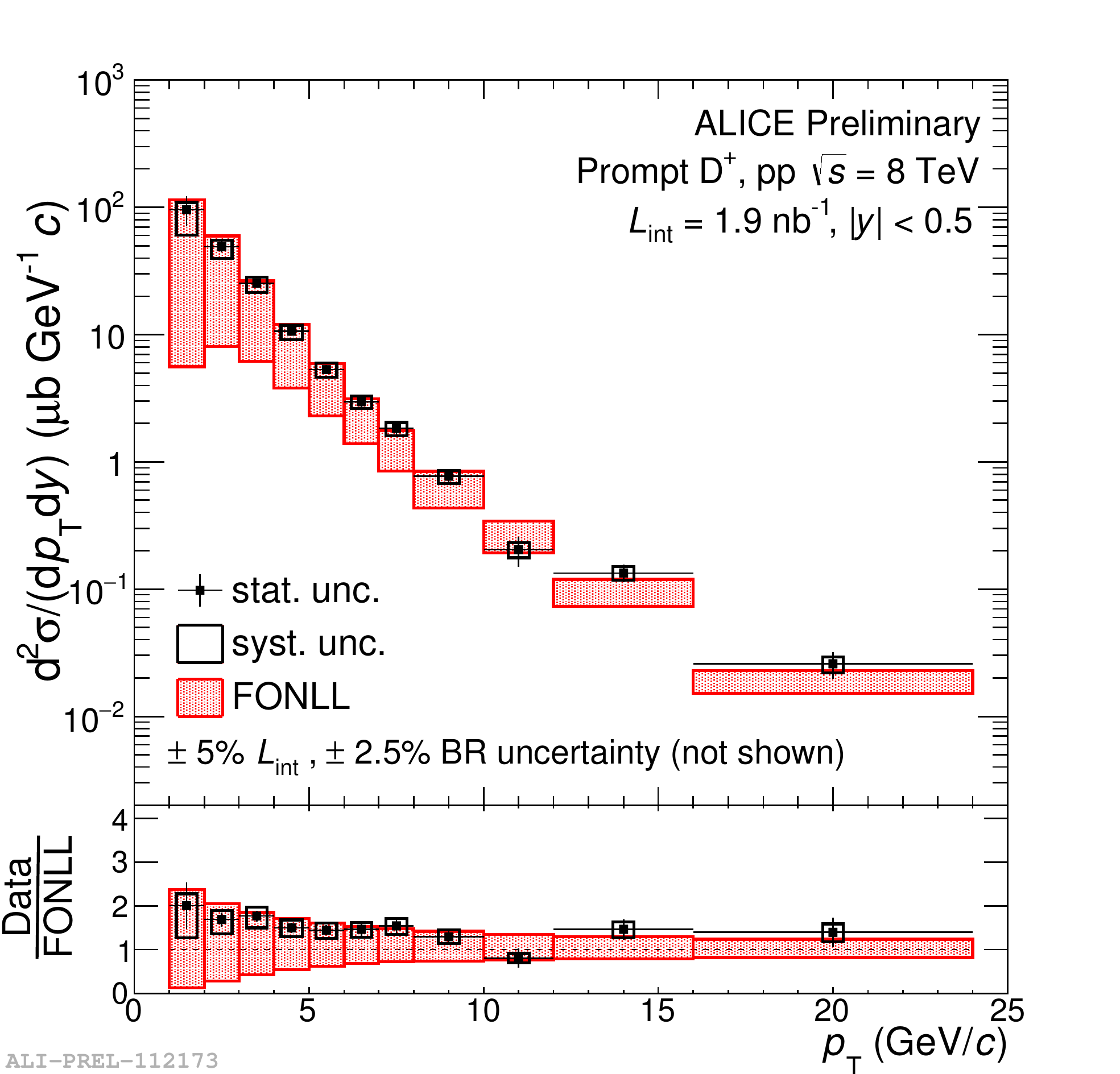}

\vspace{2cm}

\includegraphics[width=5cm]{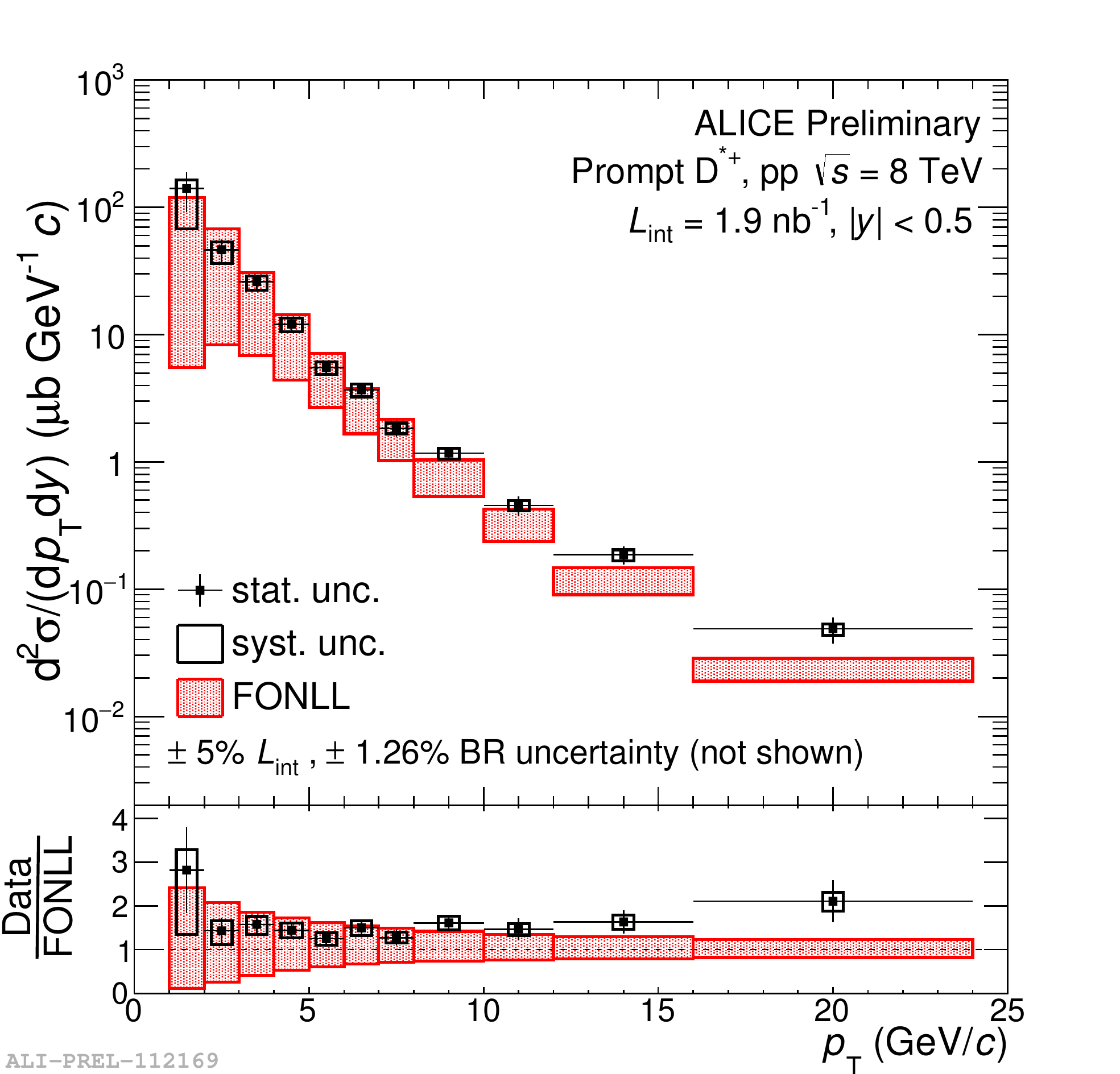}
\end{multicols}
\caption{\textit{p}$\rm_{T}$-differential cross sections for $\rm D^{+}$ and $\rm D^{*+} $ in pp collisons at  $\sqrt{s}$ = \newline8 TeV compared with FONLL theoretical predictions. Bottom: the ratio of the measured cross section and the central FONLL calculations.}
\end{figure}
These measurements were also compared with the same measurements at $\sqrt{s}$ = 7 TeV shown in Fig 2. From these figure it is clear that within the statistical fluctuations the 8 and 7 TeV results are compatible, also the ratio of the cross sections at the two energies is compatible with FONLL.
\begin{figure}[h!]
\centering
\begin{multicols}{2}
\includegraphics[width=5.5cm]{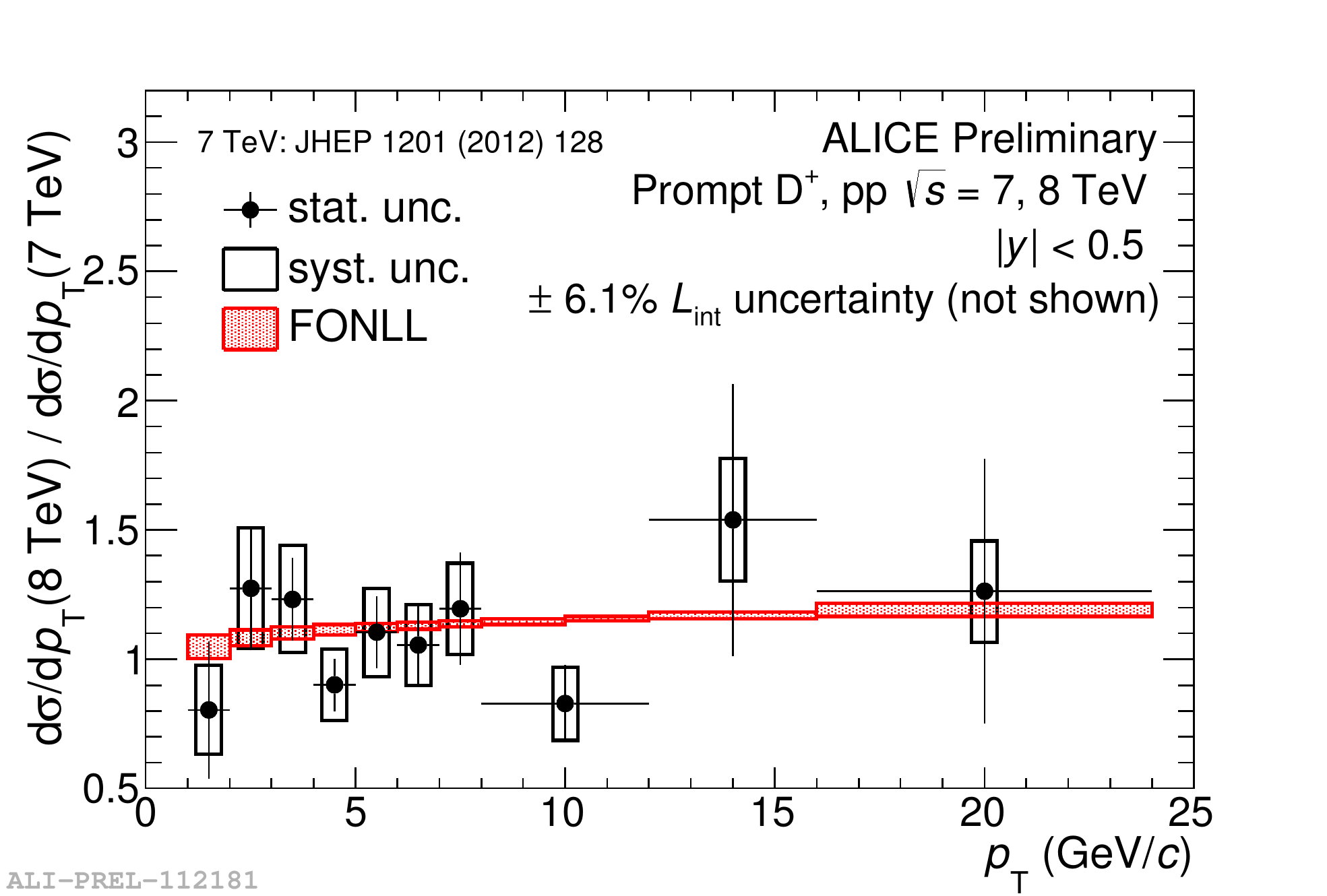}

\includegraphics[width=5.5cm]{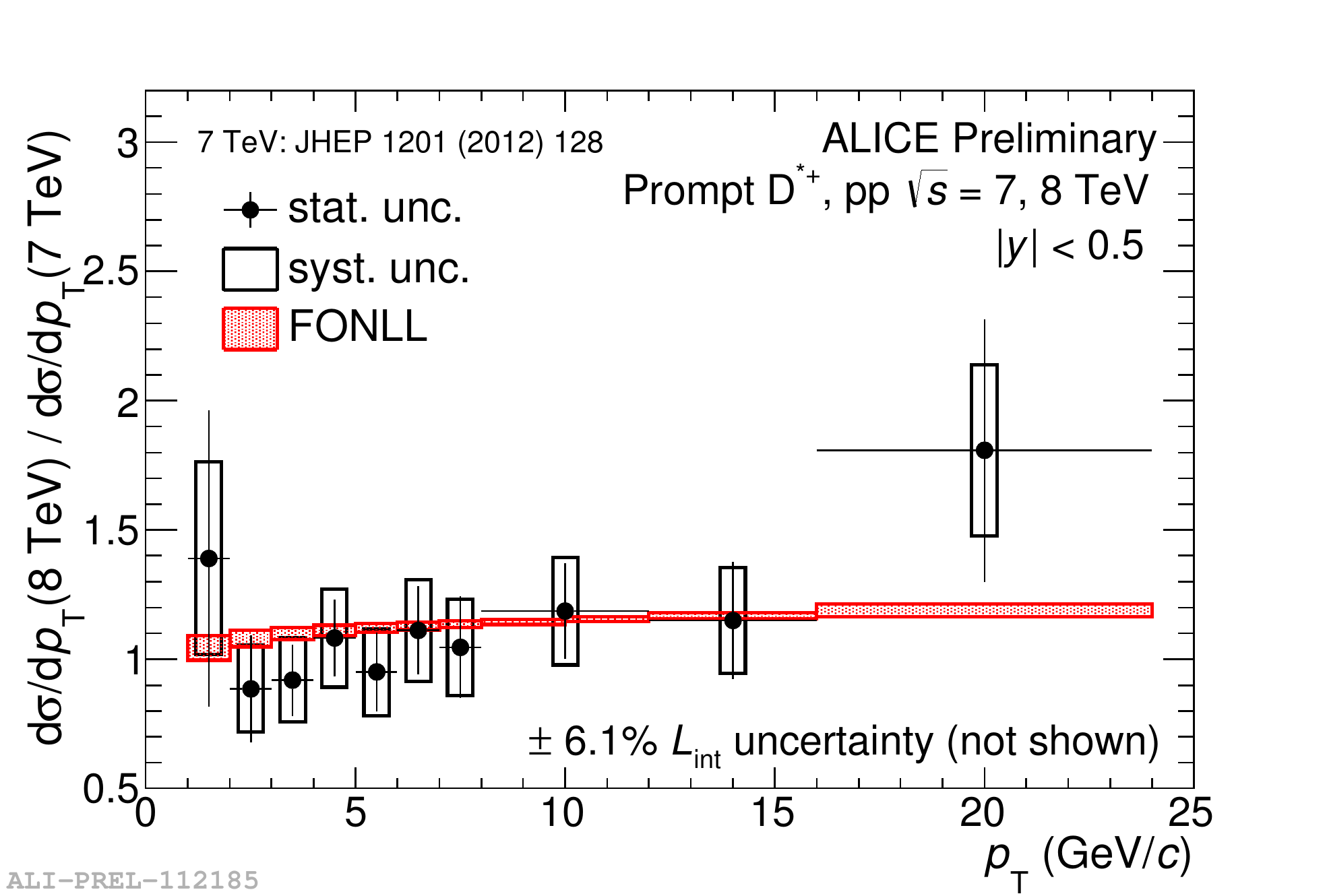}
\end{multicols}
\caption{Ratio of \textit{p}$\rm_{T}$-differential cross sections for $\rm D^{+}$ and $\rm D^{*+}$ at $\sqrt{s}$ = 7 and \newline8 TeV.}
\end{figure}

\section{D-meson yielda as a function of charged particle multiplicity}
The measurement of heavy-flavour production in pp collisions as a function of the charged-particle multiplicity produced in the collision could provide insight into the processes occurring in the collision at the partonic level and the interplay between the hard and soft mechanisms in particle production. The analysis results are presented as a function of the relative charged-particle multiplicity at central rapidity measured in pp collisions at  $\sqrt{s}$ = 7 TeV with at least one charged particle in $|\eta| < 1.0$. The results of the $\rm D^{0}$, $\rm D^{+}$ and $\rm D^{*+}$ relative yields for each \textit{p}$\rm_{T}$ interval are presented in Fig 3 as a function of the relative charged-particle multiplicity in various \textit{p}$\rm_{T}$ bins~\cite{ref:Mult}. From the figure one can observe that charm hadron yields increase with the charged particle multiplicity at central rapidity with a faster than linear increase at large multiplicities. The enhancement is qualitatively described by models including multi-parton interactions, like PYTHIA~\cite{ref:Pythia}, EPOS~\cite{ref:EPOS} and the percolation model~\cite{ref:Percolation}. 

\begin{figure}
\centering
\includegraphics[width=5.4cm]{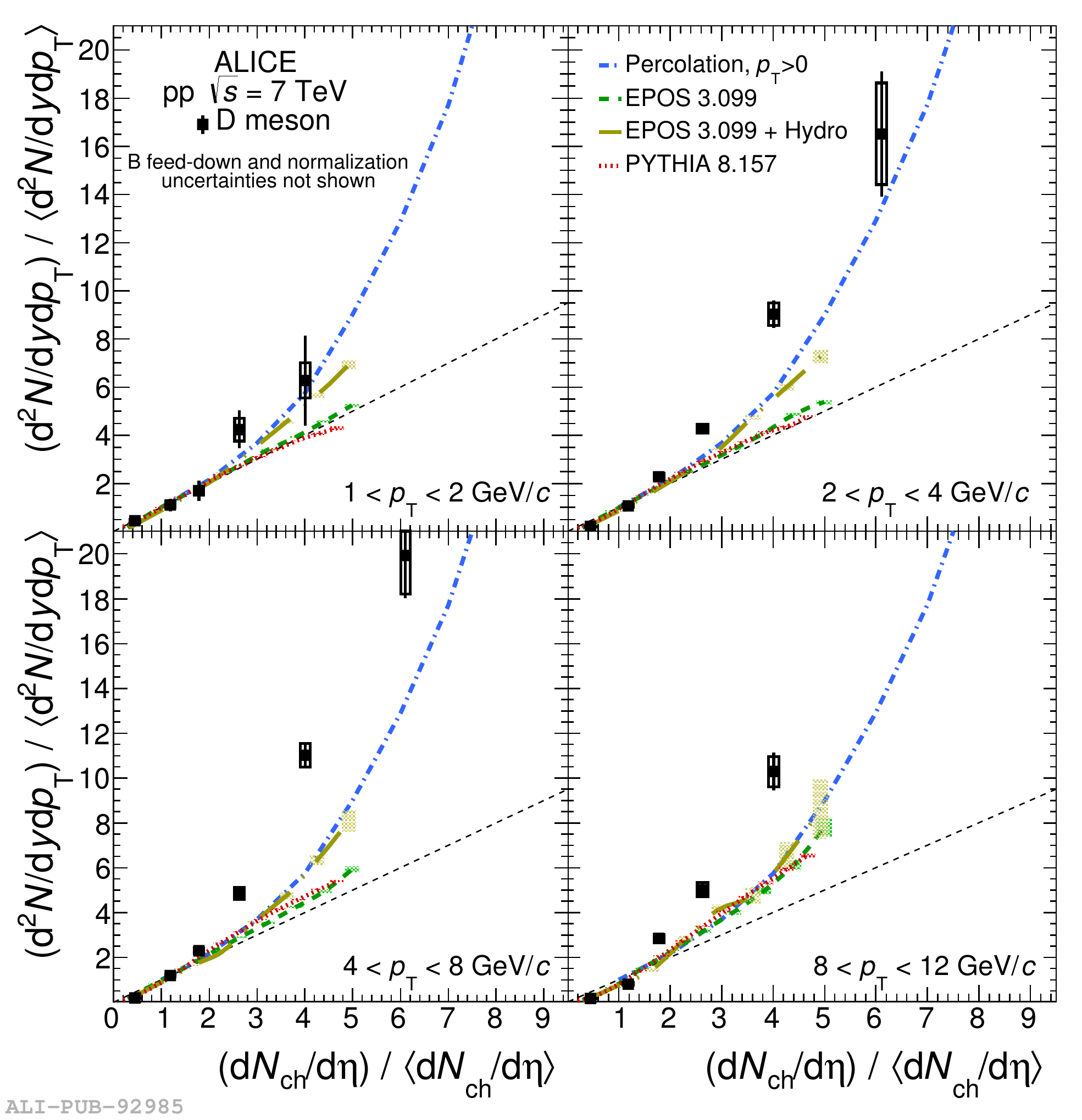}
\caption{Average D-meson relative yields as a function of the relative charged-particle multiplicity at central
rapidity in different \textit{p}$\rm_{T}$ intervals~\cite{ref:Mult}.}
\end{figure}


%
%

\end{document}